\documentclass[preprint,3p,authoryear,twocolumn]{elsarticle}
\usepackage{graphicx}
\usepackage{amssymb}
\usepackage{lineno}
\usepackage{tabularx}

\journal{Journal of Kinematics and Physics of Celestial Bodies}

\begin{document}

\begin{frontmatter}

\title{INFLUENCE OF THE EARTH'S ATMOSPHERE ROTATION ON THE SPECTRUM OF ACOUSTIC-GRAVITY WAVES}

\author{O. K. Cheremnykh}
\ead{oleg.cheremnykh@gmail.com}
\author{S. O. Cheremnykh}
\ead{ikdchereremnykh@gmail.com}
\author{D.I. Vlasov$^*$}
\ead{dima.i.vlasov@gmail.com}
\cortext[cor1]{Corresponding author.}

\address{Space Research Institute, prosp. Akad. Glushkova 40, build. 4/1, 03187, Kyiv, Ukraine}

\begin{abstract}
It was shown in a recent paper by Cheremnykh et al. (Advances in Space Research, 2022) that taking into account the rotation of the Earth's atmosphere leads to the appearance of a new region of evanescent waves with a continuous frequency spectrum on the diagnostic diagram of acoustic-gravity waves. The region is located below the lower limit of gravity waves, which is equal $2\Omega$ at all wavelengths, where $\Omega$ is the angular frequency of the atmosphere rotation. This result was obtained for high-latitude regions of the atmosphere, in which we can limit ourselves to considering only the vertical component of the Earth's rotation frequency. In the proposed paper we show that taking into account both components of the vector of the atmosphere rotation frequency $\overrightarrow{\Omega}$ - horizontal, $\Omega\cos\varphi$ , and vertical, $\Omega\sin\varphi$, where $\varphi$ is the Earth`s latitude, the dominant role in the acoustic-gravity waves propagation is played by the vertical component. It is shown that the horizontal component leads to an insignificant modification on the diagnostic diagram of the boundaries of the areas of acoustic and gravity waves, which can be neglected. Also shown that the strongest vertical component of the frequency affects the lower limit of gravity waves, which depends at all wavelengths on the latitude of the observation site and is equal to $2\Omega\sin\varphi$.
\end{abstract}

\begin{keyword}
Acoustic-gravity waves \sep acoustic waves \sep gravity waves \sep evanescent waves\sep atmosphere rotation \sep diagnostic diagram 
\end{keyword}

\end{frontmatter}

\section{INTRODUCTION}

Interest in the study of acoustic-gravity waves (AGW) in the Earth's atmosphere has not diminished for many decades [1 - 4]. In contrast to waves propagating in the ionosphere and magnetosphere, the existence of which significantly depends on the state of near-Earth plasma [5], the geometry of the Earth's magnetic field [6], external space sources [7], AGW are realized in a weakly ionized medium, and at their consideration it is possible to neglect the influence of charged particles and magnetic field. Among the sources of AGW in the Earth's atmosphere and ionosphere are earthquakes, volcanic eruptions, tornadoes, thunderstorms, solar eclipses, the movement of the solar terminator, the eruption of charged particles and the dissipation of currents in the polar regions, magnetospheric storms, substorms, strong ground explosions and missile launches[8, 9]. Martin [10] and Hines [1] were the first who pointed out the important role of AGW for many processes in the Earth's atmosphere and ionosphere. The results of numerous theoretical and experimental studies have shown that AGW make a significant contribution to the dynamics and energy of the atmosphere, providing effective interaction between different altitude levels. These waves have a significant effect on the formation of atmospheric convection and turbulence. They also affect the formation of weather systems and other atmospheric processes. The energy and momentum that AGWs transport from the bottom up into the atmosphere and ionosphere are equivalent to and even higher than those they receive from the solar wind and other sources [11, 12]. These waves also play an important role in the dynamics of ionospheric plasma [13]. In particular, having reached the ionosphere, AGWs affect the quality factor of the magnetospheric resonator for electromagnetic waves [14]. Despite the fact that the main characteristics of AGW have been studied and presented in a number of books [15 - 20] and reviews [8, 10, 21], and some progress has been made in constructing a linear [22 - 25] and nonlinear theory of these waves [26 - 30] interest in this problem is still high. At present, ground and satellite methods for observing the Earth's atmosphere and the Sun for AGW research are being improved, theoretical models are being complicated, numerical calculation programs are being improved, and problems of taking into account nonlinear effects are being considered. Until now, research on the linear theory of atmospheric perturbations has not lost relevance [4, 23, 31 - 33].

The classical theory of acoustic-gravity waves assumes the existence of waves in the isothermal stratified atmosphere that propagate freely at an angle to the horizon and evanescent wave modes that propagate horizontally [15]. For evanescent waves, the vertical component of the wave vector is a complex quantity and determines the exponential change of their amplitudes depending on the height. It is well known that the spectrum of AGWs propagating freely is continuous and on the diagnostic diagram (frequency - horizontal wave vector) [34] consists of a high-frequency acoustic region and a low-frequency region of internal gravity waves [1, 8]. The role of evanescent waves in the dynamics and energy of atmospheric processes has been studied to a much lesser extent. Until recently, only a few solutions were known for evanescent modes: the Brunt-V\"{a}is\"{a}l\"{a} oscillation ([32]), the Lamb wave [35, 36], the {\it f}-mode [37], and the $\gamma$-mode [38]. However, in [39] it was shown that the spectrum of evanescent acoustic-gravity waves is continuous. The found spectrum of evanescent acoustic-gravity waves fills the entire ``forbidden" area between free acoustic and internal gravity waves on the diagnostic diagram.

Thus, taking into account this result, the AGW spectrum on the entire plane of the diagnostic diagram is continuous and consists of acoustic and gravity regions separated by the region of evanescent waves.

A recent paper [40] considered the effect of the Earth's atmosphere on the spectrum of acoustic-gravity waves. The results of this work can be conveniently explained using the diagnostic diagram presented in Fig.1. The spectrum of acoustic-gravity waves presented in this figure consists of acoustic and gravity regions, as well as two regions of evanescent waves with a continuous spectrum. The frequencies in this figure are normalized to the frequency of the acoustic cutoff $\omega_a$, and the wave vectors are normalized to the atmospheric scale height. One region of evanescent acoustic-gravity waves (with frequencies $\hat{\omega}_1$) is located between the regions of free acoustic and gravity waves and contains a continuous set of frequencies above the frequency $2\Omega$ (Coriolis parameter). The second region of evanescent waves (with frequencies $\hat{\omega}_2$) is realized due to the rotation of the Earth's atmosphere and lies below the frequency $2 \Omega$, which at all wavelengths is the lower limit of the region of gravity waves.

The results presented in Fig. 1 were obtained for high-latitude ($\varphi\leq\pi/2$ ) regions of the Earth's atmosphere, which recently have been intensively studied [23, 25, 41, 42]. When constructing models of acoustic-gravity waves in these areas, we can restrict ourselves to taking into account only the vertical component of the Earth's rotation speed. In this paper, we will focus on studying the influence of both the horizontal and vertical components of the Earth's rotational speed on the spectrum of acoustic-gravity waves. This will allow us to study the behavior of these waves in the equatorial and mid-latitude regions. The study will be conducted within the model used in [40, 43].

Following the assumptions of this model, we neglect the curvature of the Earth's atmosphere and use the approximation of the air layer with plane-parallel boundaries. We consider the frequency of Earth rotation as a constant within the considered layer. Since, taking into account the horizontal component of the Earth's rotational speed, the atmosphere cannot be considered isotropic in the horizontal plane, we consider two cases of wave propagation. In the first case, we believe that the waves propagate in latitudinal direction, and in the second - in longitudinal direction. We restrict the consideration to free acoustic and gravity waves, because (see Fig. 1) the limiting frequencies of these waves coincide with the limiting frequencies of evanescent waves (see more in [40]). Therefore, in the diagnostic diagram, the limiting frequencies of free acoustic and gravity waves also determine the boundaries of the regions of evanescent waves.

It should be noted that the results of [40] are qualitative in nature, as they are obtained for values $\Omega$ that differ from the real ones. We focus on values $\Omega$ close to the actual frequency of rotation of the Earth's atmosphere. Since this frequency is much lower than the characteristic frequencies of acoustic and gravity waves in the Earth's atmosphere, we are primarily interested in the behavior of wave perturbations at different latitudes at the frequencies near $\Omega$ where we expect a noticeable deformation of the spectrum of acoustic and gravity waves. Note that acoustic-gravity waves with frequencies about of the Earth's rotational frequency lie in the same frequency range as the well-known Rossby waves [44, 45]. The latter play an important role in the dynamics of the Earth's atmosphere and ocean [23, 46]. Therefore, the study of acoustic-gravity waves with frequencies about of frequency of the Earth's rotation is an urgent geophysical problem.

\begin{figure}[h!]
\centering
\includegraphics[width=\columnwidth]{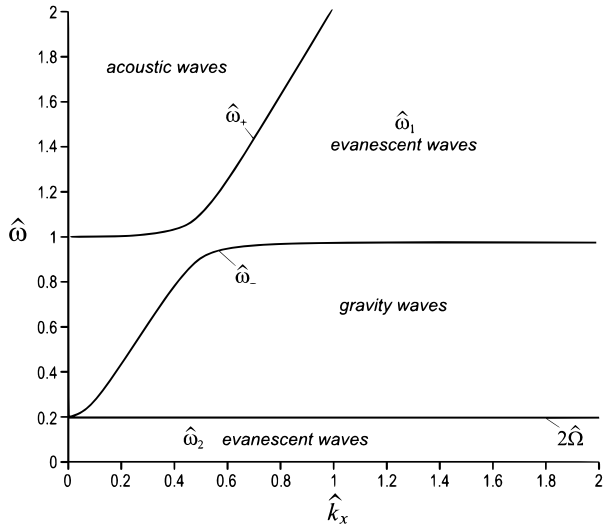}
\caption{Diagram of acoustic-gravity waves for high-latitude regions of the Earth's rotating atmosphere. Here $\hat{\omega} = {\omega}/{\omega_a}, \hat{\Omega} = {\Omega}/{\omega_a}, \hat{k}_x = k_x H, \omega_a = {C_s}/{2H}.$ }
\label{Fig.1}
\end{figure}

\section{INITIAL EQUATIONS}

We proceed from the well-known equations of dynamics of an ideal gas in the gravity field, to consider wave processes in the rotating Earth atmosphere:

\begin{equation}\label{1}
\begin{array}{l}
\frac{d\rho}{dt} + \rho {\rm{div} \overrightarrow {v}} = 0,
\end{array}
\end{equation}

\begin{equation}\label{2}
\begin{array}{l}
\frac{dp}{dt} = C_s^2 \frac{d\rho}{dt},
\end{array}
\end{equation}

\begin{equation}\label{3}
\begin{array}{l}
\frac{d \overrightarrow {v}}{dt} = -\frac{\nabla p}{\rho} - 2 \left[ {\overrightarrow {\Omega} \times \overrightarrow {v}} \right] + \overrightarrow {g},
\end{array}
\end{equation}

\noindent where

\[
\frac{d}{dt} = \frac{\partial}{\partial t} + \left( {\overrightarrow {v} \cdot \nabla} \right). 
\]

Here $\overrightarrow{v}$ is the hydrodynamic velocity of the elementary volume relative to the rotating atmosphere, $p$ and $\rho$ are the pressure and density of the gaseous medium, $\overrightarrow{g}$ is the acceleration of gravity, $C_s^2$ is the square of the speed of sound at constant entropy. Equations (1) - (3) are written in a coordinate system rotating with the Earth's atmosphere at an angular velocity $\overrightarrow{\Omega}$ [47].

We consider wave disturbances with wavelength much less than the Earth's radius. Therefore, we neglect the curvature of the Earth's atmosphere and, following [40, 43], we study the propagation of acoustic-gravity waves on the plane tangent to a spherical gas layer at some point $A$. We introduce a Cartesian coordinate system $(x,y,z)$ on this plane so that the axis $z$  is directed outward from the plane, so that $\overrightarrow{g}=-g\overrightarrow{e}_z$, the axis $x$ is directed eastward, and the axis $y$ - northward. If the latitude of the considered point is $\varphi$, then $\Omega_x=0$, $\Omega_y=\Omega\cos\varphi$, $\Omega_z=\Omega\sin\varphi$.   From now on we will assume $\overrightarrow{\Omega}=\Omega_y\overrightarrow{e}_y+\Omega_z\overrightarrow{e}_z$.

As in works [39, 40], we restrict ourselves to considering perturbations in an isothermal atmosphere, which in an unperturbed state is assumed to be in a state of static equilibrium $(\overrightarrow{v}=0)$. In such an atmosphere, the equilibrium density and pressure appearing in equations (1) - (3) are functions of $z$ and satisfy the conditions of hydrostatic equilibrium and barometric stratification

\begin{equation}\label{4}
\begin{array}{l}
\frac{dp}{dz} = -\rho g, \qquad
H = \frac{C_s^2}{\gamma g},\\\\
\frac{p(z)}{p(0)} = \frac{\rho (z)}{\rho (0)}= e^{-z/H},
\end{array}
\end{equation}

\noindent where $H$ - atmospheric scale height, $\gamma$ - adiabatic exponent.

After linearizing equations (1) - (3) with respect to the equilibrium state, differentiating the resulting equations in time, and eliminating the perturbed density and pressure, we obtain the following system of equations for small oscillations

\begin{equation}\label{5}
\begin{array}{l}
\rho \frac{\partial^2 v_x}{\partial t^2} + 2 \rho \Omega_y \frac{\partial v_z}{\partial t} - 2 \rho \Omega_z \frac{\partial v_y}{\partial t} = \\\\ \qquad = \frac{\partial}{\partial x} \left( {\rho C_s^2 {\rm{div} \overrightarrow{v}}} \right) - \rho g \frac{\partial v_z}{\partial x},
\end{array}
\end{equation}

\begin{equation}\label{6}
\begin{array}{l}
\rho \frac{\partial^2 v_y}{\partial t^2} + 2 \rho \Omega_z \frac{\partial v_x}{\partial t} = \\\\ \qquad = \frac{\partial}{\partial y} \left( {\rho C_s^2 {\rm{div} \overrightarrow{v}}} \right) - \rho g \frac{\partial v_z}{\partial y},
\end{array}
\end{equation}

\begin{equation}\label{7}
\begin{array}{l}
\rho \frac{\partial^2 v_z}{\partial t^2} - 2 \rho \Omega_y \frac{\partial v_x}{\partial t} = \\\\ \qquad = \frac{\partial}{\partial z} \left( {\rho C_s^2 {\rm{div} \overrightarrow{v}}} \right) + \rho g {\rm{div}_{\perp}} \overrightarrow{v}.
\end{array}
\end{equation}

\noindent Here $\overrightarrow{v}$ is the perturbed velocity. The rest of the designations are as follows:

\[
\nabla_{\perp} = \overrightarrow{e_x} \frac{\partial}{\partial x} + \overrightarrow{e_y} \frac{\partial}{\partial y},
\]

\[
\nabla = \nabla_{\perp} + \overrightarrow{e_z} \frac{\partial}{\partial z},
\]

\[
{\rm{div}_{\perp} \overrightarrow{v}} = \frac{\partial v_x}{\partial x} + \frac{\partial v_y}{\partial y},
\]

\[
{\rm{div} \overrightarrow{v}} = {\rm{div}_{\perp} }\overrightarrow{v} + \frac{\partial v_z}{\partial z}.
\]

Equations (5) - (7) are the initial ones for further analysis. If we neglect in them the frequency of rotation of the Earth's atmosphere $(\Omega_y = \Omega_z = 0)$ and the dependence of the perturbed quantities on the coordinate $y$ $(\partial / \partial y = 0)$, then we obtain the equations

\[
\rho \frac{\partial^2 v_x}{\partial t^2} = \frac{\partial}{\partial x} \left( {\rho C_s^2 {\rm{div} \overrightarrow{v}}} \right) - \rho g \frac{\partial v_z}{\partial x},
\]

\[
\rho \frac{\partial^2 v_z}{\partial t^2} = \frac{\partial}{\partial z} \left( {\rho C_s^2 {\rm{div} \overrightarrow{v}}} \right) + \rho g \frac{\partial v_x}{\partial x}.
\]

\noindent which are usually used to consider both acoustic and gravity waves [1, 48], as well as evanescent waves [32, 38, 39].

Note the horizontal anisotropy of wave propagation, namely, the dependence of the form of equations (5) - (7) on the direction of wave propagation. Therefore, below we will consider free acoustic and gravity waves for two cases: meridional waves $(\partial / \partial x = 0)$ and latitudinal waves $(\partial / \partial y = 0)$. In addition, we separately consider the case $(\partial / \partial x = \partial / \partial y = 0)$ for which, as will be shown below, the solution qualitatively differs from the solutions for meridional and latitudinal waves, and the frequencies corresponding to these solutions lie in the same range as the frequencies of the acoustic-gravity waves and Rossby waves.

\section{INERTIAL HORIZONTAL AND INTERNAL VERTICAL VIBRATIONS}

Let us consider the inertial motion of atmospheric gas in the horizontal plane $(x,y)$ under the action of only the Coriolis force. Such a movement, according to (5), (6), is realized under the condition $\partial / \partial x = \partial / \partial y = 0$. We seek solutions in the form

\begin{equation}\label{8}
\begin{array}{l}
v_{x,y,z} = v_{x,y,z} \left( {t , z} \right).
\end{array}
\end{equation}

The choice of perturbed velocities in the form (8) is dictated by the structure of equations (5) - (7). Substitution of (8) into (5) - (7) leads to the equations

\begin{equation}\label{9}
\begin{array}{l}
\frac{\partial v_y}{\partial t} + 2 \Omega_z v_x = 0,
\end{array}
\end{equation}

\begin{equation}\label{10}
\begin{array}{l}
\frac{\partial^2 v_x}{\partial t^2} + 4 \Omega_z^2 v_x + 2 \Omega_y \frac{\partial v_z}{\partial t} = 0,
\end{array}
\end{equation}

\begin{equation}\label{11}
\begin{array}{l}
\frac{\partial^2 v_z}{\partial t^2} - 2 \Omega_y \frac{\partial v_x}{\partial t} = \frac{1}{\rho} \frac{\partial}{\partial z} \left( {\rho C_s^2 \frac{\partial v_z}{\partial z}} \right).
\end{array}
\end{equation}

\noindent If we put $\Omega_y = 0$ in these equations, then we get the result of work [40]: horizontal and vertical oscillations occur independently. Moreover, the movement of particles of atmospheric gas in inertial oscillations will differ from their movement in internal vertical oscillations.

Assuming

\[
v_{x,y,z} \left( {t,z} \right) \sim {\rm{exp}} \left( {i \omega t + \frac{\delta z}{H}} \right),
\]

\noindent where the frequency $\omega$ is a real quantity, and the parameter $\delta$ can be a complex quantity, from (9) - (11), after simple transformations, we obtain a system of algebraic equations for $v_x$ and $v_z$:

\begin{equation}\label{12}
\begin{array}{l}
\left( {\omega^2 - 4 \Omega_z^2} \right) v_x - 2 i \omega \Omega_y v_z = 0,\\\\
\left( {\omega^2 - N^2} \right) v_z + 2 i \omega \Omega_y v_x = 0,
\end{array}
\end{equation}

\noindent where

\[
N^2 = \frac{\gamma g}{H} \delta \left( {1 - \delta} \right).
\]

It is easy to see that the frequency $N$ will be real and describes the vertical oscillations at $\Omega_y = 0 \left( {\varphi \approx \pi / 2} \right)$ in the case $0 < \delta < 1$. This frequency reaches its maximum value at $\delta = 1 / 2$ and is equal to the acoustic cutoff frequency $\omega_a = C_s / 2H$. In the particular case $\delta = 1 / \gamma$, it coincides with the Brunt-V\"{a}is\"{a}l\"{a} frequency $\omega_{BV}^2 = \left( {\gamma - 1} \right) g^2 / C_s^2$. If, following the approach of Hines [1], we put $\delta = 1/2 - ik_z H$, then the frequency $N$ coincides with the frequency of vertical acoustic waves $N^2 = \omega_a^2 + k_z^2 C_s^2$. Therefore, the value $\delta=1/2 \left( {k_z = 0} \right)$ determines the cutoff frequency of evanescent waves. Equation (12) leads to the following dispersion equation

\begin{equation}\label{13}
\begin{array}{l}
\left( {\omega^2 - N^2} \right) \left( {\omega^2 - 4 \Omega^2 \sin^2 \varphi} \right) =\\\\ \qquad = 4 \omega^2 \Omega^2 \cos^2 \varphi.
\end{array}
\end{equation}

From (13) it follows that if $\delta = 0$ or $\delta = 1$ (e.g. ${N^2 = 0}$), then oscillations with one natural frequency are realized

\begin{equation}\label{14}
\begin{array}{l}
\omega^2 = 4 \Omega^2.
\end{array}
\end{equation}

For the values $\delta$ lying in the interval $0 < \delta < 1$ in the real Earth's atmosphere, the inequality $N \sim \omega_a \gg \Omega$ is valid, in this case from (13) we find the following eigenfrequencies

\begin{equation}\label{15}
\begin{array}{l}
\omega_1^2 = \frac{4 \Omega^2 \sin^2 \varphi}{1 + \frac{4 \Omega^2 \cos^2 \varphi}{N^2}},\\\\
\omega_2^2 = \frac{N^2}{1 - \frac{4 \Omega^2 \cos^2 \varphi}{N^2}}.
\end{array}
\end{equation}

\noindent Since $\Omega / N \ll 1$ , then the estimates $\omega_1^2 \sim 4 \Omega^2 \sin^2 \varphi$, $\omega_2^2 \sim N^2$ are valid.

In the case of $\varphi \approx \pi /2 \left( {\Omega_y \approx 0} \right)$, as noted above, the horizontal oscillations are \textbf{\lq{}}split off\textbf\rq{} from the vertical ones. Equation (9) and the first equation in (12) in this case describe the periodic motion of an element of the medium in a circular orbit (inertial oscillations) with frequency $2 \Omega$ (see [22, 40, 47]), and the second equation in (12) describes vertical stratified oscillations with frequencies $N$ (see (16) and [35, 49, 50]). If expressed $\delta$ in terms of $N$ and substituted into the expression for $v_z$, then the latter exactly coincides with the Lamb solution given in [51].

Inertial oscillations in the Earth's atmosphere have practically not been studied. At the same time, they are ubiquitous in the ocean environment, in which they play an extremely important role. Measurements of disturbances in the ocean, especially from satellites [52], most often indicate the predominance of inertial oscillations, which are realized in the form of solid-body rotation of water layers. It is possible that their severity is due to the fact that the implementation of other types of disturbances is hindered by the presence of environmental stratification. As for the vertical oscillations in the Earth's atmosphere, they are the subject of careful research, including taking into account the non-isothermality of the atmosphere [50].

In the near-equatorial regions of the atmosphere $\left( {\varphi \approx 0, \Omega_z \approx 0} \right)$, oscillations are realized (see (16)) with frequencies

\begin{equation}\label{16}
\begin{array}{l}
\omega^2 \approx N^2 + 4 \Omega^2.
\end{array}
\end{equation}

Note that the frequencies $N$ appearing in equations (15) and (16) change synchronously with a change in the parameter $\delta$. Numerical calculations show that the deformation of the spectrum at frequencies $\omega \sim N \sim \Omega$ can be neglected.

\section{MERIDIONAL ACOUSTIC GRAVITY WAVES}

To consider meridional acoustic-gravity waves, we neglect the dependence on the coordinate $x$ and, following [1], seek the solution of equations (5) - (7) in the form of a harmonic plane wave

\begin{equation}\label{17}
\begin{array}{l}
v_{x,y,z} \sim {\rm{exp}} \left( {\frac{z}{2 H} - i k_y y - i k_z z} \right).
\end{array}
\end{equation}

From (5) - (7) and (17) we obtain the equations

\begin{equation}\label{18}
\begin{array}{l}
\left( {\omega^2 - k_y^2 C_s^2 - 4 \Omega_z^2} \right) v_y = \biggl[ {k_y k_z C_s^2 -} \\\\ {- 4 \Omega_y \Omega_z + i \left( {\frac{k_y C_s^2}{2 H} - k_y g} \right)} \biggr] v_z,
\end{array}
\end{equation}

\begin{equation}\label{19}
\begin{array}{l}
\\\\
\left( {\omega^2 - \frac{C_s^2}{4 H^2} - k_z^2 C_s^2 - 4 \Omega_y^2} \right) v_z = \\\\ = \biggl[ {k_y k_z C_s^2 - 4 \Omega_y \Omega_z -} \\\\ {-  i \left( {\frac{k_y C_s^2}{2 H} - k_y g} \right)} \biggr] v_y.
\end{array}
\end{equation}

Using these equations, it is not difficult to obtain the dispersion equation

\begin{equation}\label{20}
\begin{array}{l}
\omega^4 - \biggl[ {\Bigl( {k_y^2 + k_z^2} \Bigr) C_s^2 + \frac{C_s^2}{4 H^2} + 4 \Bigl( {\Omega_y^2 +}} \\\\ {{+ \Omega_z^2} \Bigr)} \biggr] \omega^2 + 4 \Omega_z^2 \left( {k_z^2 C_s^2 + \frac{C_s^2}{4 H^2}} \right) +\\\\ + 4 k_y \Omega_y C_s^2 \Bigl( {k_y \Omega_y + 2 k_z \Omega_z} \Bigr) + \\\\ + k_y^2 g^2 \left( {\gamma -1} \right) = 0.
\end{array}
\end{equation}

\noindent If we put $\Omega_y = 0 \left( {\varphi \cong \pi / 2} \right)$ in (20), then, up to replacement $k_y \rightarrow k_x$, it coincides with equation (21) obtained in [40].

Let us analyze the dependence of the frequency on the parameters $k_y$, $k_z$, $C_s$, $H$, $\Omega_y$, $\Omega_z$, $g$, and $\gamma$, writing equation (20) in dimensionless variables

\begin{equation}\label{21}
\begin{array}{l}
\hat{\omega}^4 - \biggl[ {1 + 4 \left( {\hat{k}_y^2 + \hat{k}_z^2} \right) + 4 \hat{\Omega}^2} \biggr] \hat{\omega}^2 + \\\\ + 4 \hat{\Omega}^2 \sin^2 \varphi \left( {1 + 4 \hat{k}_z^2} \right) + 16 \hat{k}_y^2 \frac{\left( {\gamma -1} \right)}{\gamma^2} + \\\\ + 16 \hat{\Omega} \cos \varphi \hat{k}_y \Bigl( {\hat{k}_y \hat{\Omega} \cos \varphi +} \\\\ {+ 2 \hat{k}_z \hat{\Omega} \sin \varphi} \Bigr) = 0.
\end{array}
\end{equation}

\noindent Here

\[
\hat{\omega} = \frac{\omega}{\omega_a},\qquad
\hat{\Omega} = \frac{\Omega}{\omega_a},\qquad
\hat{k}_y = k_y H,
\]

\[
\hat{k}_z = k_z H,\qquad
\omega_a = \frac{C_s}{2H}.
\]

\noindent In what follows $\hat{\Omega} \ll 1$, and we assume that it corresponds to the real situation in the Earth's atmosphere.

For $\hat{k}_y \rightarrow 0$, equation (21) gives the following frequencies (natural frequencies)

\begin{equation}\label{22}
\begin{array}{l}
\hat{\omega}_+^2 \cong 1 + 4 \left( {\hat{k}_z^2 + \hat{\Omega}^2 \cos^2 \varphi} \right),\\\\
\hat{\omega}_-^2 \cong 4 \hat{\Omega}^2 \sin^2 \varphi.
\end{array}
\end{equation}

\noindent Frequencies $\hat{\omega}_+$ here and below correspond to acoustic waves, and frequencies $\hat{\omega}_-$ - to gravity ones. It follows from (22) that the horizontal component of the rotational frequency modifies acoustic waves, and the vertical component modifies gravity waves. The rotation of the Earth's atmosphere in the long-wavelength limit produces the strongest effect on acoustic waves at the equator $\varphi \cong 0$, and on gravity waves at the poles $\varphi \cong \varphi = \pi / 2$. Since $\hat{\Omega} \ll 1$, then from equation (22) it follows that only the modification of the  frequencies of gravity waves can be taken into account.

In the case $\hat{k}_y \approx 1 \gg \hat{\Omega}$, equation (21) has two roots

\begin{equation}\label{23}
\begin{array}{l}
\hat {\omega}_{\pm}^2 = \frac{1}{2} \Biggl[ {1 + 4 \left( {\hat{k}_y^2 + \hat{k}_z^2} \right) \pm \Biggl( { \biggl[ {1 +}}} \\\\ {{{{+ 4 \left( {\hat{k}_y^2 + \hat{k}_z^2} \right)} \biggr]^2} - \frac{16 \hat{k}_y^2 \left[ {\gamma -1} \right]}{\gamma^2}} \Biggr)^{\frac{1}{2}}} \Biggr].
\end{array}
\end{equation}

Frequencies (23) up to replacement $\hat{k}_y \rightarrow \hat{k}_x$ coincide with the frequencies of acoustic $\left( {\hat{\omega}_+} \right)$ and gravity $\left( {\hat{\omega}_-} \right)$ waves obtained in the case $\hat{\Omega} = 0$ (see [1,3,48]). Putting $\hat{k}_z = 0$ in equation (23), we obtain expressions for the boundary frequencies of these waves (see [4,39]):

\begin{equation}\label{24}
\begin{array}{l}
\hat {\omega}_\pm^2 = \frac{1}{2} \Biggl[ {1 + 4 \hat{k}_y^2 \pm \biggl( {\left[ {1 - 4 \hat{k}_y^2} \right]^2 + }} \\\\ {{ + \frac{16 \hat{k}_y^2 \left[ {2 - \gamma} \right]^2}{\gamma^2}} \biggr)^{\frac{1}{2}}} \Biggr].
\end{array}
\end{equation}

Consider now the waves with $\hat{k_z} \gg \hat{k_y},~\hat{\Omega}, ~1$. In this case, equation (21) leads to the following frequencies:

\begin{equation}\label{25}
\begin{array}{l}
\hat{\omega}_+^2 \cong 4 \hat{k}_z^2,
\hat{\omega}_-^2 \cong 4 \hat{\Omega}^2 \sin^2 \varphi.
\end{array}
\end{equation}

\noindent It is easy to see that, under the restrictions made on $\hat{k_y}$ and on $\hat{k_z}$, the frequencies (22) coincide with the frequencies (25). It also follows from (22) and (25) that the frequencies $\hat{\omega}^2 \cong 4 \hat{\Omega}^2 \sin^2 \varphi$ determine the {\it{minimum}} frequencies of gravity waves at all wavelengths, and the frequencies $\hat{\omega}_-^2$ from (24) determine the maximum frequencies. The minimum frequencies of acoustic waves $\left( {\hat{\omega}_+} \right)$ at low values $\hat{k}_y$ are determined by equation (22), and at finite values $\hat{k}_y \approx 1$ - by equation (23). The frequencies of acoustic waves at $\hat{k}_z \gg 1$ have the form $\hat{\omega}_+^2 \cong 4 \hat{k}_z^2$.

Figures 2 show the numerical solutions of equation (21) for $\hat{k}_x = 0$, $\hat{k}_y \neq 0$ and $\hat{\Omega} = 10^{-2} $ different values of the angle $\varphi$. In fig. 2a shows the dependences $\hat{\omega}$ on $\hat{k}_y$ at $\varphi = 0$ for the finite values $\hat{k}_y$ and different values $\hat{k}_z$. It is seen that the curves with $\hat{k}_z = 0$ determine the boundary frequencies of the regions of acoustic and gravity waves. Numerical calculations show that at finite values $\hat{k}_y$ these frequencies practically do not change for different values $\varphi$ and differ little from the boundary frequencies of acoustic-gravity waves at $\hat{\Omega} = 0$. This behavior of the curves in Fig. 2a are in complete agreement with Eq. (23). The remaining figures show the behavior of frequencies near $\hat{\omega} \sim \hat{\Omega}$.

In fig. 2b shows the frequencies $\hat{\omega} \ll 1$ at $\hat{k}_y \ll 1$ and $\varphi = 0$. Their behavior is fully consistent with Eq. (22).

The frequencies shown in Fig. 2c for the values $\varphi = \pi / 4$ indicate that the lower boundary of the gravity wave region at all values of the wavelength is the frequency $\hat{\omega} = \sqrt {2} \hat{\Omega}$, which is consistent with equations (22) and (25).

\begin{onecolumn}
\begin{figure}[b!]
\centering
\includegraphics[width=\columnwidth]{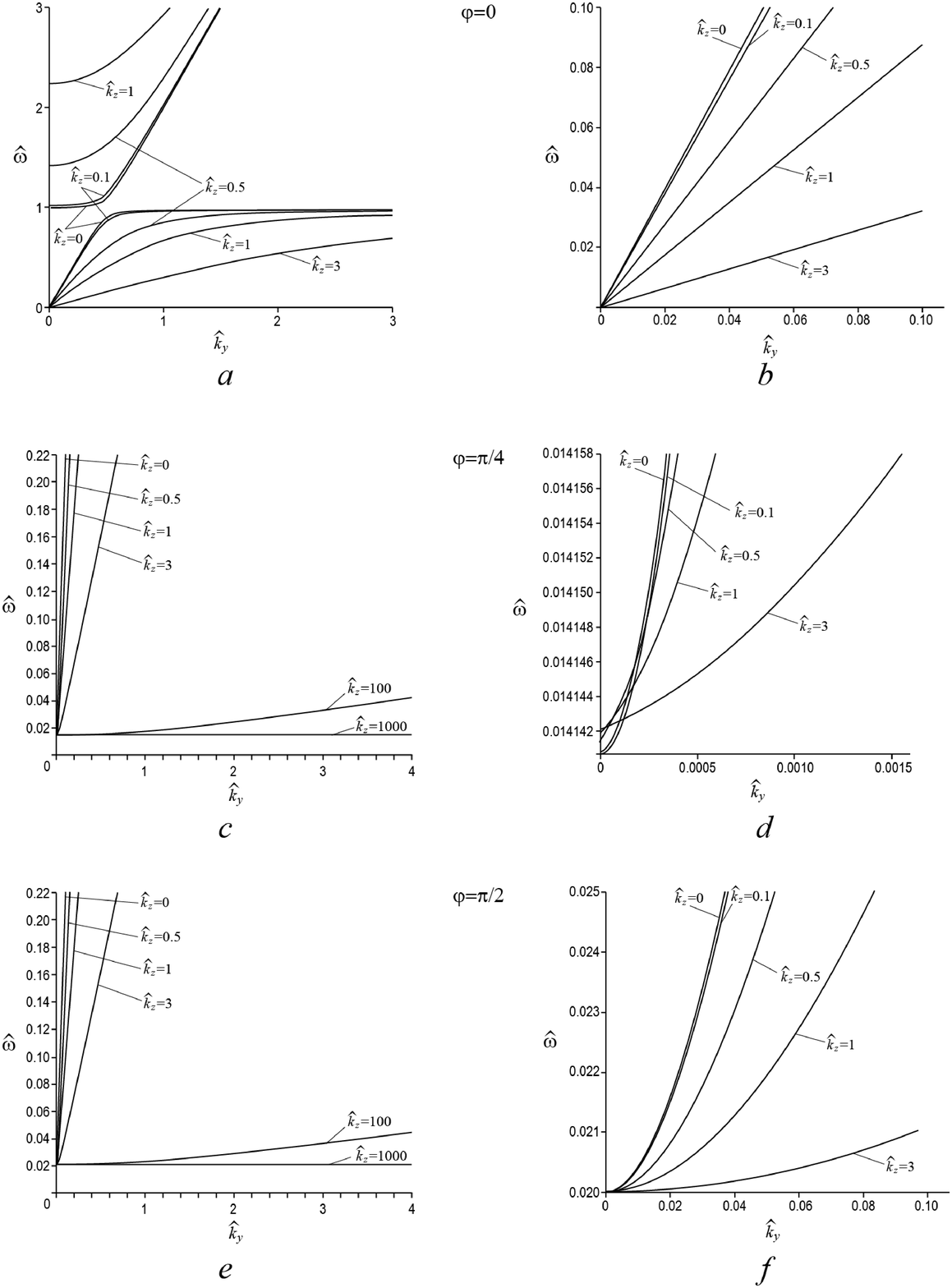}
\caption{Diagrams of acoustic-gravity waves in the rotating atmosphere of the Earth at different latitudes. Here $\hat{k}_y = k_y H, \hat{k}_z = k_z H,$ }
\label{Fig.2}
\end{figure}
\end{onecolumn}

\begin{twocolumn}

In fig. 2d shows the dependences $\hat{\omega}$ on $\hat{k}_y$ for different values $\hat{k}_z$ in the case $\hat{\omega} \ll 1$, $\hat{k}_y \ll 1$ and $\varphi = \pi / 4$. It can be seen that all frequencies at $\hat{k}_y \rightarrow 0$ are near $\hat{\omega} \cong \sqrt{2} \hat{\Omega}$, which corresponds to Eq. (22).

In fig. 2e and fig. 2f shows the frequency curves for $\varphi = \pi / 2$ and different values $\hat{k}_z$. As expected, the lower boundary of these waves is equal to $2 \hat{\Omega}$ and coincides with (25).

Thus, the influence of atmospheric rotation on meridional acoustic waves is reduced in the long-wavelength limit to an insignificant deformation of the boundaries of their regions of existence in the diagnostic diagram. As for gravity waves, for them the rotation of the atmosphere leads to the fact that the region of their existence is limited ``from below" at all wavelengths by frequency $2 \Omega \sin \varphi$. This means that the minimum frequencies of these waves will continuously change from zero at the equator to the frequency $2 \Omega$ at the pole. In work [40] for the polar regions of the atmosphere $\left( {\varphi \leq \pi / 2} \right)$ it was shown that evanescent waves are realized in the case when the frequency less than $2 \Omega$, that ``forbidden" for gravity waves, . Therefore, the frequency can be expected the $2 \Omega \sin \varphi$ will be the upper cutoff frequency for these evanescent waves. The maximum values of the frequencies of gravity waves change insignificantly due to the rotation of the atmosphere.

\section{LATITUDINAL ACOUSTIC GRAVITY WAVES}

Let us consider the waves propagating along latitude. We put $\partial / \partial y = 0$ and, following [1], we seek the solution of equations (5) - (7) in the form

\begin{equation}\label{26}
\begin{array}{l}
v_{x,y,z} \sim {\rm{exp}} \left( {\frac{z}{2H} - i k_x x - i k_z z} \right).
\end{array}
\end{equation}

\noindent Substituting (26) into (5) - (7), we obtain

\begin{equation}\label{27}
\begin{array}{l}
\left( {\omega^2 - k_x^2 C_s^2 -4 \Omega_z^2} \right) v_x = \biggl[ {k_x k_z C_s^2 +} \\\\ {+ i \Bigl({\frac{k_x C_s^2}{2H} + 2 \omega \Omega_y - k_x g} \Bigr)} \biggr] v_z,
\end{array}
\end{equation}

\begin{equation}\label{28}
\begin{array}{l}
\\\\
\left( {\omega^2 - \frac{C_s^2}{4 H^2} - k_z^2 C_s^2} \right) v_x = \biggl[ {k_x k_z C_s^2 -} \\\\ {- i \Bigl({\frac{k_x C_s^2}{2H} + 2 \omega \Omega_y - k_x g} \Bigr)} \biggr] v_x.
\end{array}
\end{equation}

From (27), (28) we find the dispersion equation

\begin{equation}\label{29}
\begin{array}{l}
\omega^4 -\biggl[ {\Bigl( {k_x^2 + k_z^2} \Bigr) C_s^2 + \frac{C_s^2}{4 H^2} + 4\Bigl( {\Omega_y^2 +}} \\\\ {{+ \Omega_z^2} \Bigr)} \biggr] \omega^2 + 4 \Omega_z^2 \left( {k_z^2 C_s^2 + \frac{C_s^2}{4 h^2}} \right) + \\\\ + k_x^2 g^2 \left( {\gamma -1} \right) + 4 \omega \Omega_y k_x g \Bigl( {1 - \frac{\gamma}{2}} \Bigr)= 0.
\end{array}
\end{equation}

\noindent Equation (29), up to replacement $k_x \rightarrow k_y$, differs from equation (21) only in the last term. In dimensionless variables, equation (29) has the form:

\begin{equation}\label{30}
\begin{array}{l}
\hat{\omega}^4 - \biggl[ {1 + 4 \left( {\hat{k}_x^2 + \hat{k}_z^2} \right) + 4 \hat{\Omega}^2} \biggr] \hat{\omega}^2 + \\\\ + 4 \hat{\Omega}^2 \sin^2 \varphi \left( {1 + 4 \hat{k}_z^2} \right) + \\\\ + \frac{16 \hat{k}_x^2 \left( {\gamma -1} \right)}{\gamma^2} + 16 \hat{\omega} \hat{k}_x \hat{\Omega} \frac{1 - \frac{\gamma}{2}}{\gamma} \cos \varphi = 0.
\end{array}
\end{equation}

\noindent Here $\hat{k}_x = k_x H$ .
	
It is easy to be convinced by direct calculations that at $\hat{k}_x \rightarrow 0$ equation (30) gives frequencies (22), and at $\hat{k}_x \sim 1 \gg \hat{\Omega}$ - frequencies (23). If we put $\hat{k}_z = 0$, then equation (30) leads to frequencies (24), which determine the minimum frequencies of acoustic waves and maximum frequencies of gravity waves. In the case $\hat{k}_z \rightarrow \infty$, equation (30) determines the frequencies (25) corresponding to ultrahigh frequencies of acoustic waves and minimum frequencies of gravity waves.

This result indicates that, despite some difference between equations (21) and (27), the natural frequencies determined by these equations practically do not differ, which is due to the smallness of the value $\hat{\Omega}$. Numerical solutions of equation (30) confirm this result.

\section{MAIN RESULTS}

It is shown that on the diagnostic diagram of the acoustic-gravity waves in the rotating atmosphere, the region of gravity waves is bounded from below by the vertical component of the Coriolis parameter $2 \Omega \sin \varphi$, where $\Omega$ is the frequency of the atmosphere rotation, and $\varphi$ is the latitude of the place. The boundary frequencies of this region, corresponding to the maximum frequencies of gravity waves, are insignificantly modified by rotation in the long-wave region.

It is found that the horizontal component $2 \Omega \cos \varphi$ of the Coriolis parameter in the diagnostic diagram leads to negligible modification of the acoustic wave boundary only in the regions of long-wave disturbances.

Earlier in [40] it was shown existence of a continuous spectra of evanescent waves that located under the lower boundary of gravity waves and between the regions of acoustic and gravity waves. It follows from the above consideration that the region of evanescent waves lying below the gravity region is absent at the equator, increases with increasing latitude and reaches its maximum value at the pole.

Note that in many works on the study of wave disturbances in the Earth's atmosphere, the so-called ``traditional" approximation is used [20]. Within the framework of this approximation, only the vertical component of the Coriolis parameter $2 \Omega \sin \varphi$ is taken into account. Such a consideration is usually justified by the fact (see [22, 26, 40]) that high-latitude regions of the atmosphere $\left( {\varphi = \pi / 2} \right)$ are considered, in which the horizontal component of the Coriolis parameter $2 \Omega \cos \varphi$ can be ignored. Our study of the influence of the rotation of the Earth's atmosphere on acoustic-gravity waves has shown that only the vertical component $2 \Omega \sin \varphi$ plays a significant role at all latitudes. The influence of the horizontal component, $2 \Omega \cos \varphi$, on the spectrum of acoustic-gravity waves can be neglected.

We analyzed the propagation of acoustic-gravity waves along latitude and along longitude and showed that the equations of small oscillations for these two cases are different. It was found that, despite some differences, both equations for values of the Earth's rotation frequency close to real lead to the same final results. This result indicates that the propagation of acoustic-gravity waves in the atmosphere with allowance for the Earth's rotation in the horizontal plane can be considered isotropic, which is consistent with the above conclusion about the need to take into account only the vertical component of the Coriolis parameter.

Inertial horizontal vibrations are also considered in the work. It is shown that the characteristic frequency of these vibrations is equal to $2 \Omega \sin \varphi$. It has been established that these vibrations are ``linked" through the horizontal Coriolis parameter , $2 \Omega \cos \varphi$, with internal vertical vibrations. The spectrum of natural frequencies of the latter is continuous and lies in the interval $0 < \omega < \omega_a$. It is shown that for real parameters of the atmosphere, the indicated ``engagement" leads to an insignificant modification of the frequencies of both types of oscillations.
\\

This work was supported by the National Research Foundation of Ukraine, project 2020.02 / 0015 ``Theoretical and experimental studies of global disturbances of natural and man-made origin in the Earth-atmosphere-ionosphere system" and the Scientific Program of the National Antarctic Center of the Ministry of Education and Science of Ukraine, as well as with partial support of the Target Comprehensive Program NAS of Ukraine for Scientific Space Research for 2018-2022.

\end{twocolumn}

\end{document}